\documentstyle[12pt,epsfig]{article}
\advance\textheight by \topskip
\begin{document}
\tolerance=100000
\thispagestyle{empty}
\setcounter{page}{0}

\def\cO#1{{\cal{O}}\left(#1\right)}
\newcommand{\be}{\begin{equation}}
\newcommand{\ee}{\end{equation}}
\newcommand{\br}{\begin{eqnarray}}
\newcommand{\er}{\end{eqnarray}}
\newcommand{\ba}{\begin{array}}
\newcommand{\ea}{\end{array}}
\newcommand{\bi}{\begin{itemize}}
\newcommand{\ei}{\end{itemize}}
\newcommand{\bn}{\begin{enumerate}}
\newcommand{\en}{\end{enumerate}}
\newcommand{\bc}{\begin{center}}
\newcommand{\ec}{\end{center}}
\newcommand{\ul}{\underline}
\newcommand{\ol}{\overline}
\newcommand{\ra}{\rightarrow}
\newcommand{\sm}{${\cal {SM}}$}
\newcommand{\as}{\alpha_s}
\newcommand{\aem}{\alpha_{em}}
\newcommand{\ycut}{y_{\mathrm{cut}}}
\newcommand{\susy}{{{SUSY}}}
\newcommand{\Dir}{\kern -6.4pt\Big{/}}
\newcommand{\Dirin}{\kern -10.4pt\Big{/}\kern 4.4pt}
\newcommand{\DDir}{\kern -10.6pt\Big{/}}
\newcommand{\DGir}{\kern -6.0pt\Big{/}}
\def\Ecm{\ifmmode{E_{\mathrm{cm}}}\else{$E_{\mathrm{cm}}$}\fi}
\def\gluino{\ifmmode{\mathaccent"7E g}\else{$\mathaccent"7E g$}\fi}
\def\photino{\ifmmode{\mathaccent"7E \gamma}\else{$\mathaccent"7E \gamma$}\fi}
\def\mgluino{\ifmmode{m_{\mathaccent"7E g}}
             \else{$m_{\mathaccent"7E g}$}\fi}
\def\taugluino{\ifmmode{\tau_{\mathaccent"7E g}}
             \else{$\tau_{\mathaccent"7E g}$}\fi}
\def\mphotino{\ifmmode{m_{\mathaccent"7E \gamma}}
             \else{$m_{\mathaccent"7E \gamma}$}\fi}
\def\ML{\ifmmode{{\mathaccent"7E M}_L}
             \else{${\mathaccent"7E M}_L$}\fi}
\def\MR{\ifmmode{{\mathaccent"7E M}_R}
             \else{${\mathaccent"7E M}_R$}\fi}
\def\lsim{\buildrel{\scriptscriptstyle <}\over{\scriptscriptstyle\sim}}
\def\gsim{\buildrel{\scriptscriptstyle >}\over{\scriptscriptstyle\sim}}
\def\jp #1 #2 #3 {{J.~Phys.} {#1} (#2) #3}
\def\pl #1 #2 #3 {{Phys.~Letta.} {#1} (#2) #3}
\def\np #1 #2 #3 {{Nucl.~Phys.} {#1} (#2) #3}
\def\zp #1 #2 #3 {{Z.~Phys.} {#1} (#2) #3}
\def\pr #1 #2 #3 {{Phys.~Rev.} {#1} (#2) #3}
\def\prep #1 #2 #3 {{Phys.~Rep.} {#1} (#2) #3}
\def\prl #1 #2 #3 {{Phys.~Rev.~Lett.} {#1} (#2) #3}
\def\mpl #1 #2 #3 {{Mod.~Phys.~Lett.} {#1} (#2) #3}
\def\rmp #1 #2 #3 {{Rev. Mod. Phys.} {#1} (#2) #3}
\def\sjnp #1 #2 #3 {{Sov. J. Nucl. Phys.} {#1} (#2) #3}
\def\cpc #1 #2 #3 {{Comp. Phys. Comm.} {#1} (#2) #3}
\def\xx #1 #2 #3 {{#1}, (#2) #3}
\def\NP(#1,#2,#3){Nucl.\ Phys.\ \issue(#1,#2,#3)}
\def\PL(#1,#2,#3){Phys.\ Lett.\ \issue(#1,#2,#3)}
\def\PRD(#1,#2,#3){Phys.\ Rev.\ D \issue(#1,#2,#3)}
\def\preprint{{preprint}}
\def\Ord{\lower .7ex\hbox{$\;\stackrel{\textstyle <}{\sim}\;$}}
\def\OOrd{\lower .7ex\hbox{$\;\stackrel{\textstyle >}{\sim}\;$}}
\def\MCH {$\tilde\chi_1^+$}
\def \CH{{\tilde\chi}^{\pm}}
\def \LSP{\tilde\chi_1^0}
\def \SNU{\tilde{\nu}}
\def \BARSNU{\tilde{\bar{\nu}}}
\def \MLSP{m_{{\tilde\chi_1}^0}}
\def \MLSP{m_{{\tilde\chi_1}^0}}
\def \N0{{\tilde\chi}^{0}}
\def \MCHMIN {\MCH^{min}}
\def \ET{\not\!\!{E_T}}
\def \LL{\tilde{l}_L}
\def \LR{\tilde{l}_R}
\def \MLL{m_{\tilde{l}_L}}
\def \MLR{m_{\tilde{l}_R}}
\def \MSNU{m_{\tilde{\nu}}}
\def \PROCESS{e^+e^- \rightarrow \tilde{\chi}^+ \tilde{\chi}^- \gamma}
\def \PI{{\pi^{\pm}}}
\def \DM{{\Delta{m}}}
\newcommand{\bQ}{\overline{Q}}
\newcommand{\ad}{\dot{\alpha }}
\newcommand{\bd}{\dot{\beta }}
\newcommand{\dd}{\dot{\delta }}
\def \CH{{\tilde\chi}^{\pm}}
\def \MCH{m_{{\tilde\chi}_1^{\pm}}}
\def \LSP{\tilde\chi_1^0}
\def \MUL{m_{\tilde{u}_L}}
\def \MUR{m_{\tilde{u}_R}}
\def \MDL{m_{\tilde{d}_L}}
\def \MDR{m_{\tilde{d}_R}}
\def \MSNU{m_{\tilde{\nu}}}
\def \MTAUL{m_{\tilde{\tau}_L}}
\def \stau1{m_{\tilde{\tau}_1}}
\def \MTAUR{m_{\tilde{\tau}_R}}
\def \mhf{m_{1/2}}
\def \MST{m_{\tilde t_1}}
\def \CHM{H^\pm}
\def \RPVC{\lambda'}
\def\tth{\tilde{t}\tilde{t}h}
\def\qqh{\tilde{q}_i \tilde{q}_i h}
\def\t1{\tilde t_1}
\def\stau1{\tilde \tau_1}
\def \MET{E{\!\!\!/}_T}  
\def \lumi{{\rm fb}^{-1}}
\def\lapp{\mathrel{\rlap{\raise.5ex\hbox{$<$}}
                    {\lower.5ex\hbox{$\sim$}}}}
\def\gapp{\mathrel{\rlap{\raise.5ex\hbox{$>$}}
                    {\lower.5ex\hbox{$\sim$}}}}
\begin{flushright}
IISc/CHEP/18/01\\
NSF-KITP-08-103
\end{flushright}
\begin{center}
{\Large \bf
Using Tau Polarization to probe the Stau Co-annihilation Region of mSUGRA Model
at the LHC
}\\[1.00
cm]
\end{center}
\begin{center}
{\large R.M. Godbole$^a$, Monoranjan Guchait$^b$   
{and} D. P. Roy$^{c,d}$}\\[0.3 cm]
{\it 
\vspace{0.2cm}

$^a$ Center for High Energy physics,\\
Indian Institute of Science,\\
Bangalore 560 012,India.
\\
\vspace{0.2cm}
$^b$Department of High Energy Physics\\
Tata Institute of Fundamental Research\\ 
Homi Bhabha Road, Mumbai-400005, India.\\
\vspace{0.2cm}
$^c$
Homi Bhabha's Centre for Science Education\\
 Tata Institute of Fundamental Research\\
V. N. Purav Marg, Mumbai-400088, India. 
\\
\vspace{0.2cm}
$^d$
AHEP group, Instituto de Fisica Corpuscular(IFIC),\\
CSIC-U.de Valencia, Correos E-46071,Valencia, Spain 
}
\end{center}

\vspace{2.cm}

\begin{abstract}
{\noindent\normalsize 
}
\end{abstract}
The mSUGRA model predicts the polarization of the tau coming from the stau to 
bino decay in the co-annihilation region to be +1. This can be exploited to 
extract this soft tau signal at LHC and also to measure the tiny mass 
differences between the stau and the bino LSP.
Moreover this strategy
will be applicable for a wider class of bino LSP models, where the
lighter stau has a right
component at least of similar size as the left.
\vspace{2cm}
\hskip1.0cm
\newpage

\section*{1.~Introduction}
\label{sec_intro}
         The minimal supersymmetric standard model (MSSM) has been the most 
popular extension of the standard model(SM) for three reasons. It provides a 
natural solution to the hierarchy problem of the SM and a natural candidate 
for the cold dark matter in terms of the lightest superparticle(LSP) along 
with the unification of gauge couplings at the GUT scale. In particular there 
is a great deal of interest in the minimal supergravity (mSUGRA) model as a 
simple and well-motivated parametrization of the MSSM. This is described
by the four and half parameters~\cite{msugra}
\br
m_{1/2}, m_0, A_0, \tan\beta~{\rm and}~{\rm sgn}(\mu),
\label{eq:one} 
\er 
the first three representing the common gaugino and scalar masses and 
trilinear coupling at the GUT scale. The $\tan\beta$ stands for the ratio of 
the two Higgs vacuum expectation values, while the last one denotes the sign 
of the mixing parameter $\mu$ between them. The magnitude of $\mu$ is fixed 
by the radiative electroweak symmetry breaking condition.

Astrophysical constraints on dark matter(DM) require the LSP to be colorless 
and neutral, while direct DM search experiments strongly disfavor sneutrino 
LSP. Thus the favored candidate for LSP in the MSSM is the lightest 
neutralino,
\br
\N0_1 = N_{11}\tilde B + N_{12}\tilde W_3 + 
N_{13}\tilde H_1 + N_{14}\tilde H_2. 
\label{eq:two}
\er
In the mSUGRA model the $\N0_1$ is dominantly bino($\tilde B$) over 
most of the allowed parameter space. Since bino has no gauge charge, it can 
pair-annihilate mainly via sfermion exchange. The large sfermion mass 
limits from LEP~\cite{pdg} makes this annihilation process inefficient, 
leading to an overabundance of DM over most of the mSUGRA parameter space.
There are essentially two narrow strips of cosmologically compatible DM 
relic density~\cite{wmap} at the opposite edges of the parameter space, 
corresponding to the large and the small $m_0$ boundaries, called the focus 
point and the stau co-annihilation regions respectively~\cite{utpal,spa}.
There is also a narrow strip in the middle called resonant annihilation
region, but only at very large $\tan\beta$($\gsim$40). Out of these only
the stau co-annihilation region is compatible with the muon anomalous
magnetic moment constraint~\cite{g2}, and also the only one which 
can be completely covered at the LHC. Therefore the stau 
co-annihilation region is a region of special interest to the SUSY 
search programme at the LHC. In particular one is looking for a distinctive
signature which will identify the SUSY signal at the LHC to this region and 
also enable us to measure the tiny mass difference $\Delta M$ between the 
co-annihilating superparticles, which is predicted to be $\sim$5\% by the
DM relic density constraint. A distinctive feature of this region is 
that a large part of the SUSY cascade decay occurs via
\br
\stau1 \to \tau \N0_1,
\label{eq:three}
\er   
leading to a $\tau$ lepton along with the canonical missing $E_T$($\MET$). 
Unfortunately, the $\tau$ leptons are very soft because of the small mass 
difference between $\stau1$ and $\N0_1$. But fortunately the $\tau$ 
polarization is predicted to be very close to +1 in the mSUGRA model. 
In this work we hope to show with the help of a generator level Monte Carlo
simulation that the positive polarization($P_\tau=+1$) of this $\tau$
signal can be exploited to extract it from the negatively polarized $\tau$
($P_\tau=-1$) background as well as the fake $\tau$ background from
hadronic jets. Moreover, the steep $p_T$ dependence of the soft $\tau$ 
signal will provide a distinctive signature for the co-annihilation region 
as well as a measure of the tiny mass difference between the co-annihilating 
particles.

In section 2 we briefly describe how to use the $\tau$ polarization. In 
section 3 we describe the stau co-annihilation region
with the help of an illustrative point in the parameter space of this region.
We also identify the main channels of SUSY cascade decay. In section 4 we 
discuss the result of the simulated SUSY signal for this point. We show how 
the polarization cut retains most of the $P_\tau=+1$ signal $\tau$-jets while
suppressing $P_\tau=-1$ background $\tau$-jets and practically 
eliminating the fake $\tau$ jets. We also estimate the significance level 
of the signal and the 
$\Delta M$ measurement from the slope of the soft $\tau$ jet signal. 
We conclude with a summary of our results in section 5. 
 
\section*{2.~Using $\tau$ Polarization}
 The best channel for $\tau$ identification
is its 1-prong hadronic decay channel, accounting for 50\% of its 
decay width. Over 90\% of this comes from 
\br
\tau \to \pi^\pm\nu(12.5\%), \rho^\pm \nu(26\%), a_1^\pm \nu(7.5\%),
\label{eq:four}
\er
where the branching fractions for $\pi$ and $\rho$ include the small K and 
$K^*$ contributions respectively, which have identical polarization 
effects~\cite{pdg}. The CM angular distribution of $\tau$ decay into $\pi$
or  a vector meson $v$($\rho,a_1$) is simply given in terms of its polarization
as,
\br
{1 \over \Gamma_\pi} {d\Gamma_\pi \over d\cos\theta} &=&
{1\over2} (1 + P_\tau \cos\theta) \nonumber \\
{1 \over \Gamma_v} {d\Gamma_{v L,T} \over d\cos\theta} &=&
{{1\over2} m^2_\tau, m^2_v \over m^2_\tau + 2m^2_v} (1 \pm P_\tau
\cos\theta)
\label{eq:five}
\er
where L,T denote the longitudinal and transverse polarization states of the 
vector meson. The fraction $x$ of the $\tau$ laboratory momentum carried by 
its decay meson i.e. the (visible) $\tau$-jet, is related to the 
angle $\theta$ via
\br
x = {1\over2} (1 + \cos\theta) + {m^2_{\pi,v} \over 2m^2_\tau} (1 -
\cos\theta),
\label{eq:six}
\er
in the collinear approximation($p_\tau \gg m_\tau$). It is clear 
from eqs.~\ref{eq:five} and \ref{eq:six} that the relatively hard part of 
the signal ($P_\tau=+1$)
$\tau$-jet comes from the $\pi$, $\rho_L$ and $a_{1\rm{L}}$ contributions; 
while for the background($P_\tau=-1$) $\tau$-jet it comes from $\rho_T$ and
$a_{1\rm{T}}$ contributions~\cite{martin}. This is the important part that 
would pass the $p_T$ threshold for $\tau$ jets.

                    Now the $\rho_T$ and  $a_{1\rm{T}}$ decays favor even 
sharing of the momentum among the decay pions, while the $\rho_L$ and 
$a_{1\rm{L}}$
decays favor uneven distributions where the charged pion carries either
very little or most of the momentum. Thus plotted as a function
of the momentum fraction carried by the charged pion 
\br
R = {p_{\pi^\pm} \over p_{\tau-jet}},
\label{eq:seven}
\er
the longitudinal $\rho$ and $a_1$ contributions peak at very low or very 
high R($\lsim 0.2$ or $\gsim 0.8$), while the transverse
contributions peak in the middle~\cite{martin,dp}.
The low R peak of the longitudinal $\rho$ and $a_1$ contributions are not 
detectable because of the minimum $p_T$ requirement on the charged track 
for $\tau$ identification ($R\ge$0.2). Now moving the R cut from 0.2 to 0.8
one cuts out the transverse $\rho$ and $a_1$ peaks
while retaining the detectable longitudinal
peak along with the single $\pi^\pm$ contribution. Thanks to the 
complementarity of these two sets of contributions, one can effectively
suppress the former while retaining most of the latter by a simple cut on the
ratio 
\br
R>0.8.
\label{eq:eight}
\er 
Thus one can suppress the hard part of the $\tau$-jet background($P_\tau=-1$), 
while retaining most of it for the signal ($P_\tau=+1$), even without 
separating the different meson contributions from one another~\cite{dp}. 
This is a simple but very powerful result particularly for the hadron 
colliders, where one cannot isolate the different meson
contributions to the $\tau$-jet in eq.~(\ref{eq:four}). This has been 
used to enhance the $P_\tau=+1$ signal of charged Higgs boson at 
Tevatron and the LHC
~\cite{dp,guchait1}. It can be also used with effect in the investigation 
of the $P_\tau=+1$ SUSY signal coming from eq.(\ref{eq:three}) at Tevatron
and the LHC~\cite{guchait2} as well as the ILC~\cite{godbole}.

         This is the first application of the $\tau$ polarization effect, 
however, in the context of probing the stau co-annihilation region.
As we shall see below the simple polarization cut of eq.~(\ref{eq:eight}) 
helps to 
suppress not only the ($P_\tau=-1$) $\tau$-jet backgrounds, but simultaneously
the fake $\tau$ jet background from hadronic jets as well. We shall
also see that the conclusion of this section remains true even
after we add the non-resonant contributions to the $\tau$ decay of 
eq.(\ref{eq:four}).  
      
\section*{3.~Stau co-annihilation region of the mSUGRA model}
For illustration we have chosen one point in the stau co-annihilation region 
of the mSUGRA model. We expect the results to hold equally for any other point 
in this region. The input mSUGRA parameters are shown in Table-1 along
with the resulting weak scale superparticle spectrum calculated using 
{\tt ISAJET}(v7.74)\cite{isasugra}. For simplicity we have taken 
$A_0=0$ and positive sign for $\mu$ since our results are not sensitive to
them. The corresponding LHC 
cross sections for the three strong processes at LO are,
\br
\sigma{(\tilde g \tilde g, \tilde g\tilde q,\tilde q\tilde q+\tilde q\tilde 
q^\ast)}=(0.46,2.4,1.3){\rm pb}.
\label{eq:nine}
\er
These are evaluated with the CTEQ3L\cite{cteq} structure functions 
setting a common 
factorization and renormalization scale at the average mass of the 
superparticle pair. But the results are insensitive to either of these 
choices. We have also checked that changing from CTEQ3L to the more
recent structure functions of CTEQ5L only change the total 
signal cross-section of eq.(\ref{eq:nine}) from 4.16 pb to 4.11 pb, 
although there are somewhat larger variations for the three individual 
processes. The largest contribution to the third process of eq.(\ref{eq:nine})
comes from qq $\to \tilde {\rm {q}}\tilde {\rm{q}}$ because of the dominance 
of the valence quark flux over gluon. Therefore, we expect uncorrelated 
production of singlet(S) and doublet(D) squarks in all
the three processes i.e. SS:SD:DD=1/4:1/2:1/4.
Thus 3/4 of the events contain one or two doublet squarks. The doublet 
squarks mainly undergo cascade decay via the $\tilde W$ dominated chargino
and neutralino states $\tilde\chi_1^\pm, \tilde\chi_2^0$,
into,
\br
\tilde\chi_1^\pm \to \stau1\nu \to \tau\nu\N0_1, \\ 
\label{eq:ten}
\tilde\chi_2^0 \to \tau^\prime \stau1 \to \tau^\prime \tau \N0_1.
\label{eq:eleven}
\er
\begin{center}
 Table 1: Masses of superparticles(in GeV), for an illustrative mSUGRA point
in the stau-coannihilation region. 
\end{center}
\begin{center}
\begin{tabular}{|c|c|c|c|c|c|c|c|c|c|c|c|c|c|}
\hline
$m_0$ & $m_{1/2}$& $\tan\beta$ & $\tilde g$&$\tilde q_L$ & $\tilde q_R$ & $\stau1$ & $\tilde\tau_2$&
$\tilde\chi_1^\pm$& $\tilde\chi_2^\pm$&$\N0_1$& $\N0_2$ & $\N0_3$& 
$\N0_4$  \\
\hline
223 &400 &40 & 942 & $\sim$884 & $\sim$856 & 172.5 & 356.8 & 307.5 & 526
 & 163 & 307 & 511.5 & 526 \\
\hline
\end{tabular}
\end{center}

Thus one expects to see one soft $\tau$ in most cascade decays, sometimes 
accompanied by a relatively hard $\tau'$. The lighter stau 
\br
\stau1 = \tilde\tau_R \sin\theta_{\tilde \tau} + \tilde\tau_L \cos\theta_
{\tilde \tau},
\label{eq:twelve}
\er 
is dominated by $\tilde\tau_R$, since there is no SU(2) contribution to 
the RGE.
The resulting polarization of the soft $\tau$ from $\stau1 \to \tau \chi $
is given by ~\cite{guchait2,nojiri},
\br
P_\tau &=& \frac{\Gamma(\tau_R) - \Gamma(\tau_L)}{\Gamma(\tau_R)+
\Gamma(\tau_L)} 
= \frac{(a^R_{11})^2 - (a^L_{11})^2}{(a^R_{11})^2 + (a^L_{11})^2}
\nonumber \\
a^R_{11} &=& - \frac{2g}{\sqrt{2}}N_{11}\tan\theta_W \sin\theta_{\tilde\tau}
- \frac{g m_\tau}{\sqrt{2} m_W \cos\beta}N_{13}\cos\theta_{\tilde\tau}
\nonumber \\
a^L_{11} &=& \frac{g}{\sqrt{2}}[ N_{12} + N_{11} \tan\theta_W]
\cos\theta_{\tilde\tau} 
-\frac{g m_\tau}{\sqrt{2} m_W \cos\beta} N_{13}\sin\theta_{\tilde\tau}.
\label{eq:thirteen}
\er
The first subscript of $a_{11}^{L,R}$ refers to the $\stau1$ and the second 
to the neutralino $\tilde\chi_1^0$. Thus the dominant term is 
$a_{11}^R \simeq - \frac{2g}{\sqrt{2}}N_{11}\tan\theta_W\sin\theta_
{\tilde \tau}$,
implying $P_\tau \simeq +1$. In fact in the mSUGRA model 
there is a cancellation between the sub-dominant terms, so that 
one gets $P_\tau\gsim$0.9 throughout the allowed parameter 
space~\cite{guchait2}. 
In the stau co-annihilation region of our interest $P_\tau\gsim$0.98. 
So one can safely set $P_\tau=$+1.

The polarization of the hard $\tau'$ from eq.(\ref{eq:eleven}) is obtained
from eq.(\ref{eq:thirteen}) replacing $a_{11}^{L,R}$ by $a_{12}^{L,R}$. The
dominant contribution comes from 
$a_{12}^L \simeq \frac{g}{\sqrt{2}}N_{22}\cos\theta_{\tilde\tau}$, implying 
$P_{\tau'} \simeq $-1. There is a similar cancellation of the sub-dominant 
contributions, leading to $P_{\tau'} \le$--0.98 in the
stau co-annihilation region. Thus one can set $P_{\tau'} $=-1.

   Finally it should be noted that probing the stau co-annihilation region 
via the $\tau$ polarization cut of eq.(\ref{eq:eight}) should be applicable 
in a wider class of MSSM with a $\tilde {\rm {B}}$ LSP. Since the coupling of 
$\tilde {\rm {B}}$ to $\tilde\tau_R$ is 2 times larger than to $\tilde\tau_L$
in eq.~(\ref{eq:thirteen}) one gets $P_\tau \gsim$ 0.6 as 
long as the $\stau1$ has a right component as least of similar size as the
left 
component~\cite{vempati}. In that case 
the R$>$0.8 cut will keep at least half of the soft tau signal.

\section*{4.~Event simulation, results and discussion}
        The SUSY signal events are generated using the event generator 
{\tt PYTHIA(v6.23)}~\cite{pythia}, which simulates superparticle 
pair production
and cascade decay for the spectrum shown in Table 1. The generated 
$\tau$ leptons are then passed through the {\tt TAUOLA}{\footnote{The current
version of TAUOLA cannot simulate $\tau$ decay when it comes from sparticle
decay. One of the authors(MG) has modified it to include these 
decay modes.}}  
package~\cite{tauola}
to simulate $\tau$ decay, which includes the effect of $\tau$ polarization 
in the hadronic decay channels of eq.(\ref{eq:four}) along with the small 
non-resonant contribution. The generated events are then passed through the
CMSJET package~\cite{cmsjet} for jet reconstruction. The jets are 
constructed using the cone algorithm in CMSJET with a cone size
of $\Delta R=$0.5. The kinematic cuts 
for jet reconstruction are $E_T>$15 GeV and $|\eta|<$4.5. Finally, the 
missing $E_T$($\MET$) is reconstructed by a vector summation of calorimetric
energies.

The events are then subject to the following selection cuts for triggering
and suppression of SM background:

\br
{\rm Number~of~jets}&\ge& 2~~{\rm with}\ \  E_T^{j_{1,2}}> 100~GeV,
|\eta_{j_{1.2}}|<4.5 \nonumber \\ 
 \MET &>& 250~GeV.
\label{eq:fourteen}
\er
From the events passing this cut we select those containing at least one 
$\tau$ jet at the generator level. We then require the softest $\tau$ jet in 
each event to satisfy
\br
15~GeV \le p_T^{\tau-jet} \le 40~GeV.
\label{eq:fifteen}
\er  
This is our sample of soft $\tau$ jet events. We try to simulate the efficiency
of tracker isolation on the $\tau$ jet by requiring that (i) it has one and 
only one charged track (leading track) of $p_T^{ltr}>$6 GeV within narrow 
signal cone of $\Delta R_s=0.1$ measured with respect to its calorimetric 
energy deposit, and (ii) there is no other charged track in a 
surrounding isolation cone of $\Delta R_I=0.4$ with $p_T^{ch}>$3(1) GeV. 
The $p_T^{ch}>$1 GeV isolation cut ensures higher purity of $\tau$-jets 
at the cost of a lower efficiency of $\tau$ identification, 
as genuine $\tau$ jets can be often accompanied by 
such a charged track in the environment of the LHC.

It should be noted here that in a full simulation the tracker isolation cut
is supplemented with calorimetric cuts for complete $\tau$ identification 
\cite{sasha}, which is beyond the scope of the present work.
Since the efficiency of these supplementary cuts is quite high ($\sim$0.8)
we shall simply assume it to be 1. 
We know from full simulation studies\cite{sasha} that $\tau$-jets 
identified via tracker isolation and calorimetric cuts match very
well(in direction and energy) with the generator level $\tau$-jets. Since
we are unable to incorporate the calorimetric cuts, however, we shall
work only with generator level $\tau$-jets, satisfying the tracker 
isolation, as in the case of ref.\cite{drees}. We shall add to these 
the fake 
$\tau$-jets coming from the generator level hadronic jets, which satisfy
the tracker isolation cut. Thus we approximately incorporate the tau 
identification efficiency and purity in our analysis.  
Working with generator level $\tau$-jets allows 
us to tag it to its 'mother'.
Thus we can separate the soft $\tau$ jet signal
of the stau decay of eq.(\ref{eq:three}) from the background coming from the
various other sources in the cascade decay; and look at the effect
of the polarization cut of eq.(\ref{eq:eight}) on each of them. This will
help us to understand the major contributors to the soft $\tau$ 
jet background. We shall also consider the largest SM background
to the soft $\tau$-jet signal, coming from $t \bar t$ and W+multijet 
channels.

\vspace{0.5cm}
Table 2: Number of events for the process $\tilde q\tilde g$
with and without tracker isolation. 
``Mothers'' of each $\tau$ jet, shown 
below the 3rd row are found by direct tagging in the event generator. 
(I) no tracker 
isolation (II) with tracker isolation $p_T^{ch}>$3~GeV and $p_T^{ltr}>6$ 
GeV (III) with tracker isolation $p_T^{ch}>$1~GeV and $p_T^{ltr}>6$~GeV.
Below the 3rd row the parenthetic entry in each block is obtained with 
the $R>$0.8 cut.
\begin{center}
\begin{tabular}{|c|c|c|c|}
\hline
Selection & I & II & II \\
\hline
No. of events simulated & 1500k & 1500K & 1500K \\
\hline
At least one $\tau$-jet &431283 &320207  & 254150\\
With 15 GeV $<p_T^{\tau-\rm{jet}}<$40~GeV & 156555 &108398  & 86137 \\
\hline
$\stau1$ & 87061 &61925 &48415 \\
  & (42259) &(35239) & (28178)\\
\hline
$\N0_2$ & 27073 &22813 & 19178\\
 & (10852) &(9475) & (7913)\\
\hline
$\N0_3$ & 644 &573 & 509\\
  & (345) &(315) & (283)\\ 
\hline
$\N0_4$ & 712 &639 & 548\\
 & (377) &(353) & (304)\\
\hline
W & 22419 &16532 & 13033\\
 & (6770) &(5358) & (4227)\\
\hline
Z & 787 &611 & 494\\
 & (320) &(265) & (212)\\
\hline
h & 2994 &2366 & 1925\\
 &(1189) & (1033) & (831)\\
\hline
$\tilde\tau_2$ & 98 &79 & 59\\
 & (40) &(37) & (30)\\
\hline
None of the above 'mothers' & 14767 &2860 & 1976\\
 & (5287) &(1265) & (939)\\
\hline
\end{tabular}
\end{center}
\vspace{0.5cm}
We have generated 1.5 million events for the $\tilde q\tilde g$ production, 
which has the largest SUSY cross-section of eq.(\ref{eq:nine}) i.e. 2.4 pb. 
Thus it corresponds to an effective luminosity of about 600~$\lumi$. This will 
ensure that the predicted events have greater statistical accuracy than the 
data, so that our error estimates are primarily controlled 
by the latter. Table 2 shows the effects of the above mentioned cuts on these
events. The 2nd row shows that about 1/3rd of the events contain
at least one $\tau$-jet, where the suppression factor 
includes the effects of the selection cut(\ref{eq:fourteen}) and 
the 1-prong hadronic branching ratio of $\tau$ lepton. It also shows the 
effect of the tracker isolation cuts in the three columns. We get an 
efficiency factor of about 2/3(1/2) corresponding to the $p_T^{ch}>$3(1) 
GeV on the accompanying
charged tracks. The 3rd row shows that about 1/3 of these events have the 
softest $\tau$ jet in the $p_T$ range of eq.(\ref{eq:fifteen}). 
The subsequent rows tag these $\tau$ jets to their 'mother' and
show the effect of the polarization cut of eq.~(\ref{eq:eight}) in each 
of these cases in parentheses.
We see that about half of these soft $\tau$ jets come from the signal
process of eq.~(\ref{eq:three}) having $P_\tau=$+1. About 60\% of them 
survive the
polarization cut. The background is dominated by the $P_\tau=$-1 
$\tau$-jets coming from $\tilde\chi_2^0$ of eq.~(\ref{eq:eleven}), and 
the W bosons
produced in cascade decay. About 30\%(40\%) of the $\tau$-jets from
W($\tilde\chi_2^0$) decay survive the polarization cut(\ref{eq:eight}).
The reason for the higher survival probability of the $\tau$ jets 
from the $\tilde\chi_2^0$ decay(\ref{eq:eleven}) is that they are
quite hard to start with. Hence even the relatively soft
part of these $\tau$-jets, coming from the $\rho_L$
and $\pi$ decay channels given in eq.~(\ref{eq:five}), survive the 
modest $p_T$ cut of
eq.(\ref{eq:fifteen}). The $\tau$ jets tagged to none of the listed 
'mothers' are expected to come mainly from b hadron decay. 
\\

Table 3: Estimation of faking efficiency for QCD and SUSY
($\tilde g \tilde g$) jet events; QCD jet events are for 
$\hat p_T$ range of 15-40 GeV, where $\hat p_T$ denotes the transeverse 
momentum in the partonic centre of mass frame.
\begin{center}
\begin{tabular}{|c|c|c|}
\hline
 Selection & $p_T^{ch}>$1 GeV, $p_T^{ltr}>$6~GeV & $p_T^{ch}>$3~GeV, 
$p_T^{ltr}>$6~GeV \\
\hline
process$\rightarrow$ & QCD \ \ \  \ \ \ SUSY & QCD~~~~~~~~~~~SUSY \\
\hline
No. of events simulated& $10^6$~~~~~~~~~~~~300K& $10^6$~~~~~~~~~~~~300K \\  
\hline
Total no. of jets & 181377~~~~~~~~~~~~349989 & 827424~~~~~~~~~~~~347303 \\
\hline
Total no. of (fake)$\tau$ jets &6486~~~~~~~~~~~~9183&123100~~~~~~~~~~~~38473\\
Total no. of (fake)$\tau$-jets with R cut & 376~~~~~~~~~~~~313& 
2496~~~~~~~~~~~~651\\
Faking efficiency & 0.034~~~~~~~~~~~~0.026&0.15~~~~~~~~~~~0.11\\
Faking efficiency with R cut&0.0015~~~~~~~~~~~0.0009&0.003~~~~~~~~~~~0.0019\\
\hline
\end{tabular} 
\end{center}
 Table 3 demonstrates the effect of polarization cut on the fake 
$\tau$ jets. It presents the fake $\tau$ jets coming from the hadronic 
jets produced in SUSY($\tilde g\tilde g$) and QCD processes without the 
selection cuts of eq.~(\ref{eq:fourteen}). It shows that the fake 
$\tau$-jets background 
from the hadronic 
jets in this SUSY cascade decay are quite large, particularly
for the tracker isolation cut of $p_T^{ch}>$3~GeV. But it gets practically
eliminated by the R cut of eq.(\ref{eq:eight}).
The faking efficiency of 11(3)\% for the isolation cut of $p_T^{ch}>$3(1) GeV 
falls to 0.2(0.1)\% after the R cut. For comparison we also show in this
table the faking efficiencies of hadronic jets for 1 million simulated QCD
events. The faking 
efficiencies before and after the R cut are seen to be very
similar to those of the SUSY events. This is evidently a very powerful
result, which is not unexpected though. A hadronic jet can fake an 
one prong $\tau$-jet by a rare fluctuation, when all but one of the 
constituent particles (mostly pions) are neutral. Then requiring
the single charged particle to carry more than 80\% of the total
jet energy requires a second fluctuation which is even rarer.
Normally identification of $\tau$ jets down to
$p_T$=15~GeV would be difficult at the LHC because of the fake backgrounds,
which is alleviated now by the $R >$0.8 cut. Note that
the R cut automatically raises the $p_T$ threshold of the leading
track to 12 GeV.
\\

\vspace{0.5cm}
Table 4: The SM backgrounds from $t \bar t$ and W+multijet processes.
In the latter case separate simulations are done for the two $\hat p_T$
ranges of W, as shown in the table. 
\begin{center}
\begin{tabular}{|c|c|c|}
\hline
Process$\to$ & $t \bar t$ & W+multijets \\
& &$\hat p_T=$100-300 GeV~~~~~~~~~300-1000GeV \\
\hline
No. of events simulated & 3$\times 10^6$ &$10^6$~~~~~~~~~~~~~~$10^6$  \\
\hline
$\MET>$250 GeV &15318 &1996~~~~~~~~~~150073 \\
Jet cut & 15318 & 1996~~~~~~~~~~150072  \\
$M_{eff}>$750 GeV & 9277 &136~~~~~~~~~~73841 \\
One $\tau$-jet &1250 & 13~~~~~~~~~~12931\\ 
$M_T>$50 GeV & 417 &2~~~~~~~~~~2651 \\
15~GeV$<p_T<$40~GeV & 150  &0~~~~~~~~~~~602 \\
From W & 148 &0~~~~~~~~~~~601\\
With R cut & (40) &0~~~~~~~~~(143) \\
\hline
Raw Cross sections(pb) & 500 & 1500~~~~~~~~30 \\
Number of events for ${\cal L}$=10~fb$^{-1}$ & & \\
Without R cut & 247 & $<1$~~~~~~~180 \\
With R cut & (67) & ($<1$)~~~~~~~(43) \\
\hline   
\end{tabular} 
\end{center}
Since the $R > $0.8 cut practically eliminates the 
fake $\tau$ jet 
background for both the isolation cuts, we choose to work with 
the $p_T^{ch}>$ 3 GeV
cut for two reasons. Firstly it is less demanding on the tracker momentum
resolution; and secondly it has a higher efficiency for identifying genuine
$\tau$ jet, as seen above. 
We have found that imposing the selection cuts of eq.\ref{eq:fourteen} 
suppresses
the QCD process very effectively in agreement with ref.\cite{drees}.
Supplementing this with the faking efficiency with R$>$0.8 cut makes the 
QCD background to the SUSY signal completely negligible. The largest
SM background comes from the $t \bar t $ and W+multijet processes. In order 
to control these backgrounds we supplement the selection cuts of 
eqs.(\ref{eq:fourteen}),(\ref{eq:fifteen}), with two more kinematic cuts,
\br
M_{eff} = \MET + E_T^{j_1} + E_T^{j_2} > 750~{\rm GeV},\ \  
M_T(\tau-jet, \MET)>50~{\rm GeV},
\label{eq:sixteen1}
\er  
where $M_T=\sqrt{2 p_T^{\tau-jet} \MET(1-\cos\phi(p_T^{\tau-jet},\MET))}$
and $\phi$ is the azimuthle angle between $p_T^{\tau-jet}$ and $\MET$. 
Table 4 summarizes the effect of all these kinematic cuts on the simulated
$t \bar t$ and W+multijet events. In both cases the $\tau$ coming from
W decay has $P_\tau$=-1. The R$>$0.8 cut is seen to further 
reduce these backgrounds to the 25\% level. The last two row show the 
number of background events for luminosity ${\cal L}$=10~fb$^{-1}$. 
Here we have
used the LO cross-sections for these SM processes for consistency 
with the SUSY signal.

\vspace{0.5cm}
Table 5: Events from three SUSY processes 
for tracker isolation, $p_T^{ch}>$3~GeV, $p_T^{ltr}>$6~GeV. In the last column
the total number of events are presented after normalizing the contributions 
from various sub-processes with their respective cross sections for 
luminosity of 10$\lumi$. As in Table 2, the parenthetic entry  
in each block is obtained with the R$>$0.8 cut.
\begin{center}
\begin{tabular}{|c|c|c|c|c|}
\hline
Process$\to$ & a & b & c & a+b+c  \\
&$\tilde g\tilde g$ &$\tilde q \tilde q+\tilde q\tilde q^\ast$ 
&$\tilde q\tilde g$&Lumi=10/fb  \\
\hline
No. of event simulated & 300K &800K &1000K &  \\
\hline
$\MET>$250 GeV & 183548 &624350  &744353  &    \\
Jet cut & 183548 & 624350 & 744352 &   \\
$M_{eff}>$750 GeV &155364 & 596914 & 685045&   \\
One $\tau$-jet & 35876 &114016 & 152093 & \\ 
$M_T>$50 GeV & 25083 & 76437 & 105108 &  \\
\hline
15~GeV$<p_T<$40~GeV & 8065 & 25618 & 34558 & \\
\hline
$\stau1$ & 3209   & 17143 & 16827 & 731  \\
 & (1811) & (9681) & (9544)  & (414)  \\
\hline
$\N0_2$ &1991 & 7043 & 8927 & 358 \\
    & (776) & (2841)  & (3651) & (146) \\
\hline
$\N0_3$ & 84 & 10 & 235 & 7 \\
 & (46) & (5) & (112) & (3) \\
\hline
$\N0_4$ & 80 & 57 & 223 & 7 \\
 & (45) & (31) & (130) & (4) \\
\hline
W & 2049 & 873  & 6084 & 191\\
 & (656) & (283) & (1937) & (61) \\
\hline
Z & 52 & 31 & 230 & 7 \\
 & (22) & (12) & (99)  &  (3)  \\
\hline
h & 219 & 396 & 861 & 30 \\
 & (94) & (153) &(343)  &  (12)  \\
\hline
$\tilde\tau_2$ & 13 & 1 &24 &$<1$ \\
 & (5) &(1) & (14) & ($<$1)\\
\hline
None of the above mothers & 368 & 64 & 1147 & 34\\
 & (166) & (30) & (508) & (15)\\
\hline
Fake as a $\tau$ jet & 1681  & 2006 & 5096 & 180 \\
With R cut & (40) & (62) & (126) & (5)\\
\hline
Total number of events:~Signal & & & & 731(414) \\
SUSY Background &&&& 814(249)  \\
SM Background & & & & 427(110)  \\
\hline
\end{tabular} 
\end{center}
Table 5 shows the simulated SUSY events for
all the three processes of eq.(\ref{eq:nine}).
The effects
of kinematic cuts on the soft $\tau$ jet events and their distributions to
the various sources are very similar for the three production processes.
The last column gives the total number of events from the sum of the three
processes normalized by their respective cross sections, for a luminosity
of 10$\lumi$. We see from Table 5 that without the R cut the total
background is almost twice as large as the signal. Imposing the R cut 
eliminates the fake $\tau$ background and reduces the total background
to below the signal size. 
\begin{figure}
\vspace{-2.0cm}
\hspace{-1.5cm}
\begin{center}
\includegraphics[width=12cm,height=10cm]{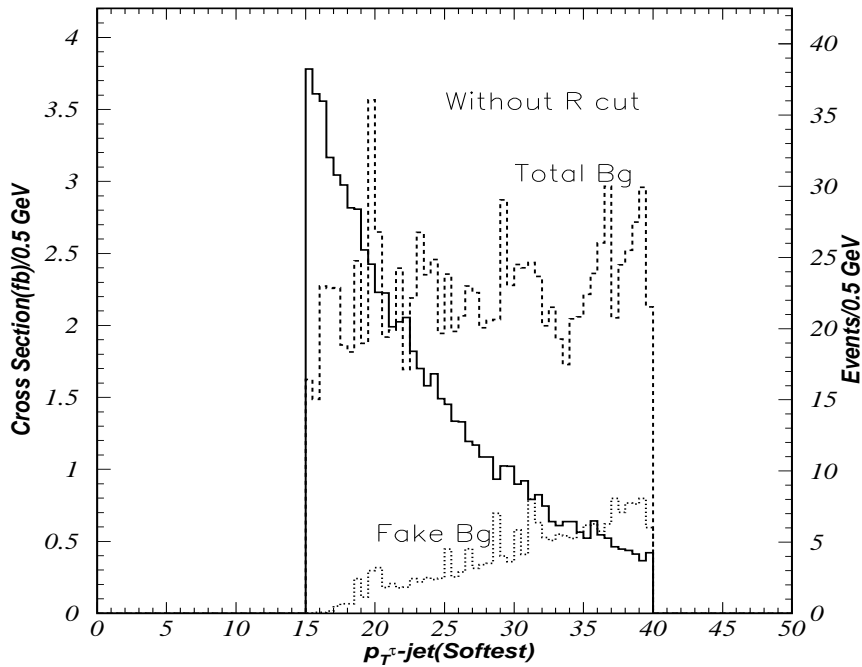}
\caption{
$p_T$(in GeV) of softest $\tau$-jet for signal(solid) and background 
processes.}
\end{center}
\end{figure}

     Fig.1 shows the $p_T$ distributions of the soft $\tau$ jet signal and
the backgrounds before the R cut, where the contribution
from the fake $\tau$ jet background is indicated 
separately by the dotted line.
The vertical scale on left gives the cross section
in fb/0.5 GeV, while that on right gives the number of events/0.5 GeV for a 
luminosity(${\cal L}$) of 10$\lumi$. 
Fig.2 gives the corresponding distributions after the R 
cut of eq.~(\ref{eq:eight}). It clearly shows a steep rise of the 
soft $\tau$ jet 
signal above
the background at the low $p_T$ end.

As we see from Tables 4 and 5, the SUSY background is larger than the SM 
background because the former is not suppressed by the selection cuts. 
Moreover while one can independently estimate the SM background, the 
SUSY background depends on the relevant SUSY masses. It will be
very hard therefore to disentangle this background from the 
SUSY signal. It is for this reason that the $\tau$ polarization 
plays a very important role in signal extraction. The $P_\tau$=+1 signal 
and the $P_\tau=-1$ background(SUSY and SM) are known a priori 
to have very distinct R dependences. Likewise one knows the distinct
R dependence of the fake $\tau$ background from hadronic jets. Thus for a 
given $p_T$ range of the $\tau$-jet events one can make use of the 
observed R dependence to separate the $P_\tau$=+1 $\tau$ signal
from the $P_\tau$=-1 $\tau$ as well as the fake $\tau$ backgrounds.
We shall assume that when real data becomes available, 
all the important SUSY masses, except the small mass
difference $\Delta M$ between $\stau1$ and $\N0_1$, can be roughly 
estimated (to $\sim$25\% accuracy, say), via mass reconstructions
using the hard jets along with the hard $\tau$-jet of $p_T>$40 GeV,
as widely discussed in the literature. 
Then one can estimate
the size of the SUSY background from this hard $\tau$-jet
region of the data and parametrize it in terms of a cascade decay fit 
to this data using these SUSY masses. One can then
extrapolate it to the $p^{\tau-jet}_T<$40 GeV region using
this parametrization. This will provide a fairly reliable estimate of the SUSY
background. 
An observed excess of low $p^{\tau-jet}_T$ events
over the (SUSY+SM) background, which has been suppressed using the 
polarization cut of 
R$>$0.8(Fig.2), will constitute the soft $\tau$ signal from eq.(\ref{eq:three}).
Thus the procedure for estimating the SUSY background from real data is quite
clear. In the absence of real data, however, we see no better alternative
to estimate the SUSY background than the simple one adopted 
here, i.e take it directly from the SUSY event generator along with the
signal for the illustrative mSUGRA point in the stau-coannihilation region of
interest. We feel it suffices for our main purpose of demonstrating 
the importance of the polarization cut in extracting the soft $\tau$ SUSY
signal from background. The importance of this cut for signal extraction
is evident from a comparison of figures 1 and 2. 
       
The steep slope of the soft $\tau$ jet signal from the $\stau1$ 
decay of eq.(\ref{eq:three}) over 
the low $p_T$ region can be used to extract the signal as well as to measure
the tiny mass difference $\Delta M$, responsible for this steep slope.
For this purpose, we divide the $p_T$ range of the soft $\tau$-jet of Fig.2 
into two parts, $p_T=$15-25 GeV and 25-40 GeV. Then we consider the ratio
of the numbers of events coming from the two parts, i.e
\br
D_s= \frac{N_{15-25}=288}{N_{25-40}=126} =2.3 
\label{eq:seventeen}
\er         
as a distinctive parameter for extracting the signal and measuring the 
mass difference $\Delta M$. Being a ratio of cross sections this quantity
should be dominated by the statistical error. 
\begin{figure}
\vspace{-2.0cm}
\hspace{-1.5cm}
\begin{center}
\includegraphics[width=12cm,height=10cm]{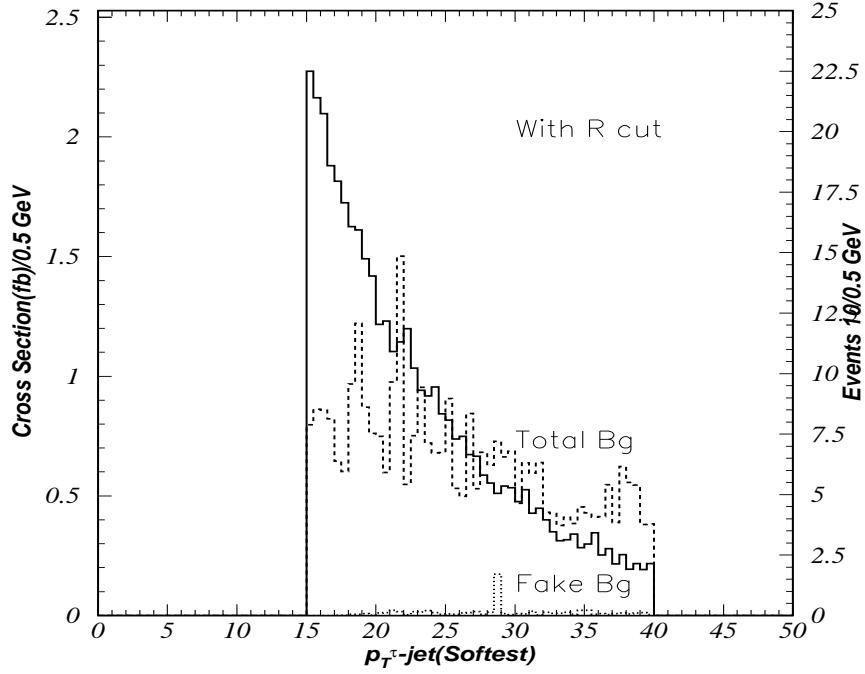}
\vspace{-0.7cm}
\caption{
Same as Fig.1, but with R cut of eq.(\ref{eq:eight}). 
}
\end{center}
\end{figure}
From Fig.2 we can estimate
this ratio for the signal+background
and for the background only. 
We get
\br
D_{S+B} &=& 467/306 = 1.52\pm 0.14 ; 
\label{eq:eighteen}  
\er
\br
D_B&=&178/180=0.98 \pm 0.14;
\label{eq:nineteen}
\er  
where the statistical errors shown are for 10$\lumi$ 
luminosity run of the LHC.
We see that this luminosity will be
enough to extract the signal at the 3$\sigma$ level. This will go up to 
$\sim$10$\sigma$ level at the 100$\lumi$
luminosity since statistical error goes down like the $\sqrt{{\cal L}}$.
\begin{figure}
\vspace{-2.0cm}
\hspace{-1.5cm}
\begin{center}
\includegraphics[width=12cm,height=10cm]{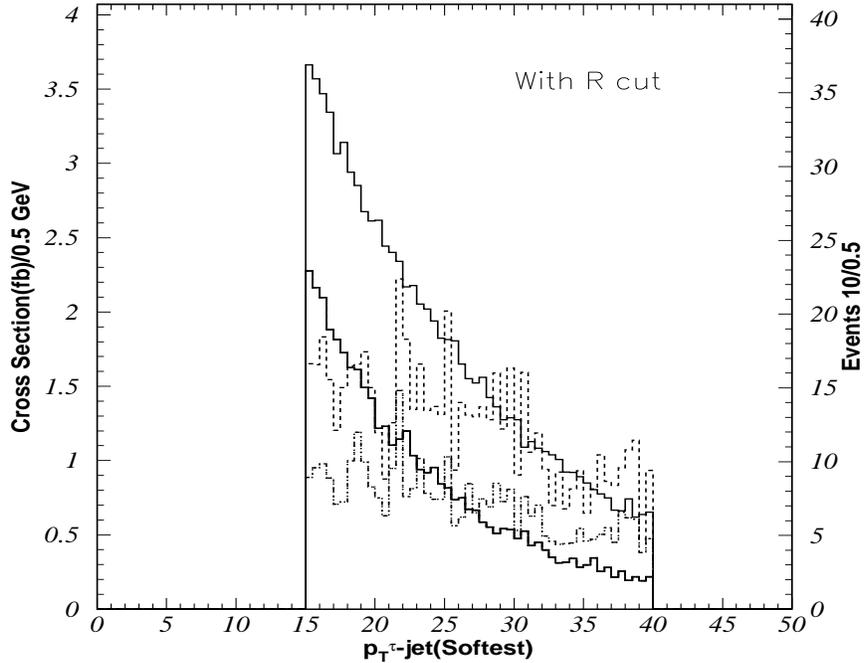}
\caption{
$p_T$(in GeV) of softest tau jet for signal(solid) and 
background(dashed) processes. The 
upper solid and dashed histograms are  
for signal and background processes for $\Delta M=$15 GeV and the lower ones
are for $\Delta M=$10 GeV. These are subject to R cut of eq.(\ref{eq:eight}).
}
\end{center}
\end{figure}

  To estimate the accuracy in the determination of the small mass
difference $\Delta M$ between $\stau1$ and $\N0_1$, we have chosen a second 
mSUGRA point near that shown in Table 1, increasing $m_0$ by 6 GeV. It 
increases $\Delta M$ from nearly 10 GeV to 15 GeV without practically
changing any other mass. Fig.3 compares the resulting soft $\tau$ jet
cross sections(signal and background) for the two points,
the RHS scale again corresponding to the number of events for a luminosity
of 10$\lumi$. One can see a flattening of the low $p_T$ slope with increasing
mass difference. The ratio(\ref{eq:eighteen}) for the upper curves is
\br
D_{(S+B)}(\Delta M=15~GeV) &=& 552/411=1.34\pm 0.10.
\label{eq:twenty}
\er
Thus one can estimate $\Delta M$ to 50\% accuracy at $\sim$1.5$\sigma$ 
level with a luminosity of 10$\lumi$. The significance level goes up to 
5$\sigma$ level for a luminosity 
of 100$\lumi$, which makes the statistical error smaller by factor 
of 3.2. 

Note that for estimating the small mass differences $\Delta M$ from the 
$p_T$ slope of the soft $\tau$-jet one has to assume some 
knowledge of the other SUSY masses. It follows from simple 
kinematics that in the rest frame of the decaying 
$\stau1$ the $p_T^{\tau-jet} \sim \Delta M$. But the boost factor
relating it to the $p_T^{\tau-jet}$ in the laboratory frame,
$p_{\stau1}$/M, depends on the other SUSY masses. As mentioned
above, we assume that these masses will be known
to $\sim$25\% accuracy, say, from the mass reconstruction 
programme. Therefore, we need to check the robustness of our
result to a $\sim$25\% variation in the SUSY masses, other than $\Delta M$.
For this purpose we have chosen a higher point on the stau-coannihilation
strip with $m_0$ and $m_{1/2}$ values $\sim$25\% higher than those
of Table 1. The SUSY mass parameter for this point are shown in Table 6. 
The resulting ratio for this point is 
\br
D_{S+B}(\Delta M=13~{\rm GeV}) = \frac{641}{444}=1.44\pm 0.11.
\label{eq:twentyone} 
\er
Note that this mass difference, $\Delta M=13$~GeV, is approximately midway
between those of the two points represented in Fig.3, $\Delta M=$ 10 GeV and 
15 GeV. So we expect the resulting ratio (\ref{eq:twentyone}) to be
nearly midway between those of eqs.(\ref{eq:eighteen}) and (\ref{eq:twenty}),
which is indeed the case. Of course there is room for a more detailed 
investigation of the sensitivity of the $\Delta M$ estimate to the other
SUSY masses, which is beyond the scope of this illustrative work.  
\begin{center}
 Table 6: Masses of superparticles (in GeV), for a higher mSUGRA point
in the stau-coannihilation region compared to that of Table 1.
\end{center}
\begin{center}
\begin{tabular}{|c|c|c|c|c|c|c|c|c|c|c|c|c|c|}
\hline
$m_0$ & $m_{1/2}$& $\tan\beta$ & $\tilde g$&$\tilde q_L$ & $\tilde q_R$ & $\stau1$ & $\tilde\tau_2$&
$\tilde\chi_1^\pm$& $\tilde\chi_2^\pm$&$\N0_1$& $\N0_2$ & $\N0_3$& 
$\N0_4$  \\
\hline
262 &500 &40 & 1156 & $\sim$1083 & $\sim$1043 & 219 & 425 & 390 & 635
 & 206 & 390 & 622 & 635 \\
\hline
\end{tabular}
\end{center}
 It should be mentioned here that we have selected a large value of 
$\tan\beta$ for our illustrative  mSUGRA point of Table 1 simply because
it gives stau-coannihilation region for comparable values 
of $m_0$ and $m_{1/2}$. But we have also checked the results for a 
low values $\tan\beta=$10. It gives stau-coannihilation region for
$m_0<<m_{1/2}$. However, the size of the soft $\tau$-jet SUSY signal and
background are similar to those of the present analysis.     
   
Finally let us note that an independent method of probing the stau 
co-annihilation region at the LHC has been investigated in~\cite{arnowitt} 
via a 
ditau-jet signal coming from $\N0_2$ decay given in eq.~(\ref{eq:eleven}). 
It contains
a hard ($p_T>$40 GeV) $\tau$ jet along with the soft one, correspodning to
the $\tau'$ and $\tau$ of eq.(\ref{eq:eleven}) respectively. The background
is suppressed by taking the difference of opposite-sign and 
same-sign ditau events. The upper edge of this di-taujet invariant mass plot 
gives an estimate of the tiny mass difference $\Delta M$, again assuming
some knowledge of the other SUSY masses. We find 
in our simulation, however, that the number of such soft and hard 
di-taujet events 
are suppressed by an order of magnitude compared to the soft taujet events
analyzed here. Therefore, the present analysis has the benefit of utilizing
a much larger fraction of the SUSY events, triggered
via the selection cut of eq.~(\ref{eq:fourteen}). But, of course, the di-taujet
channel offers an independent probe of this signal and hence should be
pursued independently. Let us point out that the $\tau$ polarization 
effect can be also exploited to improve the result of this di-taujet 
analysis. In
particular imposing the $R>$0.8 cut of
eq.(\ref{eq:eight}) on the soft $\tau$ jet and the complementarity cut on the 
hard one will ensure that both the $\tau$ jets are leading ones - i.e 
they carry most of the momentum of the respective $\tau$ leptons, with 
accompanying soft neutrinos. This will steepen the upper edge
of the di-taujet invariant mass distribution
and improve the resulting estimate of $\Delta M$, apart from
suppressing the soft $\tau$-jet background.

\section*{5.~Conclusion}
The stau co-annihilation region of the mSUGRA model is a region of special
interests to the SUSY search programme at the LHC. In particular one is looking
for a distinctive signature, which will identify the SUSY signal at the LHC 
to this region and also enable us to measure the tiny mass difference 
$\Delta M$ between the co-annihilating superparticles, which is predicted
to be $\sim$5\% by the DM relic density constraint. A distinctive feature
of this region is that a large part of the SUSY cascade decay occurs via 
$\stau1 \to \tau \N0_1$, leading to a $\tau$ lepton along with the canonical 
missing $E_T$($\MET$). Admittedly the resulting $\tau$-jets are very 
soft because of the small mass difference between $\stau1$ and $\N0_1$ states.
On the other hand its polarization ($P_\tau$) is predicted to be very close 
to +1. We have shown here with the help of generator level Monte Carlo
simulation that the positive polarization($P_\tau$=+1) of this signal can be
exploited to extract it from the negatively polarized($P_\tau$=-1) $\tau$-jet
background as well as the fake $\tau$ background from hadronic 
jets. Moreover, the steep $p_T$ dependence of the soft $\tau$-jet signal is 
shown to provide a distinctive signature for this co-annihilation region
as well as a measure of the tiny mass difference $\Delta M$ between the 
co-annihilating superparticles. The significance levels of this signal and 
$\Delta M$ measurement are estimated for the LHC luminosicties of
10$\lumi$ and 100$\lumi$.       
 
\section*{6.~Acknowledgments}

One of the authors(DPR) was supported 
in part by the BRNS(DAE) through 
Raja Ramanna Fellowship scheme and in part by MEC grants FPA 2005-01269,
SAB2005-0131. RMG and MG acknowledge support from the Indo French Centre
for Promotion of Advanced Scientific Research under project number 3004-2.
RMG wishes to thank KITP for hospitality for a 
visit during which part of the work was finished.
RMG and MG wish to thank Sabine Kraml on related issues.
MG is thankful to Gobinda Majumdar for many useful 
discussions.

\end{document}